\documentclass[prl,twocolumn,showpacs,superscriptaddress]{revtex4-1}

\usepackage{graphicx,amssymb,amsmath,psfrag,color}

\begin{document}

\title{Linear quantum quench in the Heisenberg XXZ chain: time dependent Luttinger model description of a lattice system}


\author{Frank Pollmann}
\affiliation{Max-Planck-Institut f\"ur Physik komplexer Systeme, Dresden, Germany}

\author{Masudul Haque}
\affiliation{Max-Planck-Institut f\"ur Physik komplexer Systeme, Dresden, Germany}

\author{Bal\'azs D\'ora}
\email{dora@kapica.phy.bme.hu}
\affiliation{BME-MTA Exotic  Quantum  Phases "Lend\"ulet" Research Group and Department of Physics, Budapest University of Technology and
  Economics, Budapest, Hungary}

\date{\today}

\begin{abstract}

We study variable-rate linear quenches in the anisotropic Heisenberg (XXZ) chain, starting at the XX
point.  This is equivalent to swithcing on a nearest neighbour interaction for hard-core bosons or an interaction quench
for free fermions.  The physical observables we investigate are: the energy pumped into the system during the quench,
the spin-flip correlation function, and the bipartite fluctuations of the $z$ component of the spin in a box.  We find excellent
agreement between exact numerics (infinite system time-evolving block decimation, iTEBD) and analytical results from bosonization, as a
function of the quench time, spatial coordinate and interaction strength.  This provides a stringent and much-needed
test of Luttinger liquid theory in a non-equilibrium situation.


\end{abstract}


\maketitle


While it is difficult to study genuine non-equilibrium dynamics in solid state systems due to the presence of many
relaxation channels (phonons, impurities, interactions etc.), cold atoms in optical lattices provide an ideal laboratory
for non-equilibrium investigations due to the high degree of control over various dissipation mechanisms.  Cold-atom
experiments in the past decade have explored a wide variety of non-equilibrium quantum dynamics in previously
inaccessible regimes \cite{BlochDalibardZwerger_RMP08,polkovnikovrmp}.  This has also led to an increasing amount of
theoretical activity \cite{dziarmagareview,polkovnikovrmp}.  Key issues include thermalization as well as equilibration and
their relation to integrability \cite{polkovnikovrmp}, pumping beyond the adiabatic limit or quantum fluctuation
relations \cite{rmptalkner}, and universal near-adiabatic dynamics in quantum critical systems
\cite{dziarmagareview,polkovnikovrmp}.  Linear quenches occuring over a finite time can interpolate between the more
familiar limits of an instantaneous quench and an adiabatic sweep.  Very recently, a few experiments have examined the
response of many-body experiments to such finite-time quenches \cite{Bloch2010NPhys,DeMarco2011PRL}.  It is thus of
vital current interest to address the dynamics under linear sweeps of system parameters such as interaction strength.

The response of a system to an external perturbation depends sensitively on its spatial dimension, as famously
demonstrated in the experiment of Ref.~\cite{kinoshita}. There,  one dimensional interacting bosons did not reach thermalization within the experimental timescale, 
while 
 their higher dimensional realizations did.  One dimensional systems are notoriously strongly
correlated due to the limited phase space for scattering.  The non-interacting ground state is immediately destroyed by interactions, forming a Luttinger liquid (LL) in
many instances \cite{giamarchi,nersesyan}, and described by critical phenomena of collective modes with anomalous
(non-integer) power-law dependence of correlation functions.

Quantum quenches in Luttinger liquids have been addressed by several authors
\cite{cazalillaprl,iucci,uhrig,doraquench,perfetto,karrasch,rentrop,coira}.  However, it is not clear to what extent the
Luttinger liquid (LL) picture, which is a genuine low energy description, is applicable under non-equilibrium
circumstances \cite{coira}. For an abrupt interaction change, certain observables revealed universal LL behaviour\cite{karrasch}.
 While in equilibrium, the relevance or irrelevance of a given process can be
classified (e.g. using power counting), such an approach is not reliable out of equilibrium, where additional energy
scales emerge (e.g. quench duration \cite{doraquench} and the difference between initial and final parameters).

Therefore, to understand the applicability of the continuum LL description for quenches, one needs to go beyond the LL
paradigm by either considering additional terms in the Hamiltonian (termed irrelevant in equilibrium) or by comparing
the results of the LL theory to numerical simulations on lattice models. We have undertaken the second option, and
performed extensive numerical simulations of arbitrary rate quenches on the XXZ Heisenberg model and compared these to
bosonization results. Similar approach was undertaken in Ref. \cite{karrasch} for the case of a sudden interaction quench.

The model under study is the XXZ Heisenberg model, which reads as
\begin{gather}
H=\sum_mJ\left(S_m^xS_{m+1}^x+S_m^yS_{m+1}^y\right)+J_z(t)S_m^zS_{m+1}^z
\label{xxz}
\end{gather}
where $m$ indexes the lattice sites with lattice constant set to unity, and $J>0$ is the antiferromagnetic exchange
interaction.  We are going to manipulate $J_z$ as a function of time as $J_z(t)=J_zQ(t)$, with $Q(t)$ encoding the
explicit quench protocol, switched on at $t=0$.  We concentrate on a linear quench, namely with $Q(t<\tau)=t/\tau$ and
$Q(t>\tau)=1$.  Via a Jordan-Wigner transformation \cite{EPAPS}, the XXZ Heisenberg Hamiltonian maps onto spinless 1D
fermions with nearest neighbour interaction \cite{giamarchi}:
\begin{gather}
H=\sum_m \frac{J}{2} \left(c^+_{m+1}c_m +\textmd{h.c.}\right)+J_z n_{m+1}n_m,
\end{gather}
up to an irrelevant shift of the energy.  The $c$'s are fermionic operators. 
Alternatively,  $S_l^+$ acts as a hard core boson creation operator to site $l$, and the model maps to the hopping problem of
hard core bosons, interacting with nearest-neighbour repulsion.
 This can conveniently be treated using Abelian bosonization,
after going to the continuum limit as \cite{giamarchi,nersesyan}
\begin{equation}
H=\sum_{q\neq 0} \omega(q)b_q^\dagger b_q
+\frac{g(q,t)}{2}[b_qb_{-q}+b_q^+b_{-q}^+],
\label{hamilton}
\end{equation}
with $\omega(q)=v|q|$ ($v=J$ being the bare "sound velocity"),
and $b_q^\dagger$ the creation operator of a bosonic density wave, and $g(q,t)=g_2(q)|q|Q(t)$, $g_2(q)=g_2\exp(-R_0|q|/2)$ with $R_0$ the range of the interaction.
The connection between the two models is established as $-1\ll {g_2}/{2v}={J_z}/{\pi J} \ll 1$.
The velocity  renormalization \cite{giamarchi} by $J_z$ is neglected since it does not affect the physics we discuss to leading order.

We describe  time-evolution  using the Heisenberg equation of motion, leading to \cite{doraquench}
\begin{gather}
b_q(t)= u_q(t)\;b_q(0) +v_q^*(t)\;b^+_{-q}(0)\;,
\label{timedepop}
\end{gather}
where all the time dependence is carried by the time dependent Bogoliubov coefficients $u_q(t)$ and $v_q(t)$, and the operators on the r.h.s.\ refer to non-interacting bosons before the quench.
The bosonic nature of the quasiparticles requires $|u_q(t)|^2-|v_q(t)|^2=1$.
The Bogoliubov coefficients are determined from \cite{doraquench}
\begin{gather}
i\partial_t\left[\begin{array}{c}
u_q(t)\\
v_q(t)\end{array}\right]=\left[\begin{array}{cc}
\omega(q) & g(q,t)\\
-g(q,t) & -\omega(q)
\end{array}\right]
\left[\begin{array}{c}
u_q(t)\\
v_q(t)\end{array}\right],
\label{beq}
\end{gather}
with the initial condition $u_q(0)=1$, $v_q(0)=0$.

We now obtain various dynamical quantities using the bosonization approach and compare the results to numerical
simulation of the quench on the lattice system of Eq. \eqref{xxz}.  The numerical simulations were performed using a
combination of a matrix-product state (MPS) \cite{Fannes-1992} based infinite density matrix renormalization (iDMRG)
\cite{White-1992,Kjaell-2012,McCulloch-2007} and the infinite time evolving block decimation (iTEBD)  \cite{Vidal-2007}
algorithms \cite{,dmrgmps,EPAPS}.  In our implementation of the two algorithms we use infinite, translationally invariant
systems. Working in the limit of infinite systems has the advantage that no finite size effects show up and the only
approximation is the finite bond dimension ($\chi$) of the matrix-product state (MPS).  In critical systems (as the one
we are studying), the finiteness of $\chi$ induces a finite correlation length $\xi\propto\chi^{\kappa}$ with $\kappa$ being a model specific parameter
\cite{Pollmann-2009,Tagliacozzo-2008}.  For our simulations, we use MPS's with bond dimensions of up to $\chi=2000$ to
ensure that the induced correlation length does not affect our results. We first use the iDMRG method to find the ground
state by optimizing variationally a wavefunction in the MPS representation. Then the actual quench is simulated using
the iTEBD technique. This technique is based on a Suzuki-Trotter decomposition of the time-evolution operator and
provides an efficient algorithm to perform the real-time evolution of the MPS during the quench. We choose a time-step
of $\delta t = 0.01 J^{-1}$ and use a second-order Trotter decomposition.

\begin{figure}[th]
\psfrag{x}[t][][1][0]{$J\tau$}
\psfrag{y}[b][t][1][0]{$(\langle H\rangle-E_{gs})\tau^2J^3/NJ_z^2$}
\includegraphics[width=5.5cm]{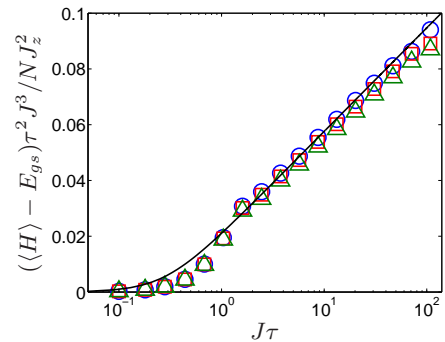}
\caption{(Color online) The heating is plotted from iTEBD for $J_z/J=0.1$ (blue circles), 0.2 (red squares) and 0.4
  (green triangles) together with the prediction of Eq. \eqref{excessenergy} (black solid line), using
  $R_0=2v\tau_0=0.5622$ (from Fig. \ref{fluctbox}, $J_z/J=0.1$ data).  The agreement remains excellent for small
  variations of $R_0$ as well.}
\label{heating}
\end{figure}

\begin{figure}[th]
\psfrag{x}[t][][1][0]{$l$}
\psfrag{y}[b][t][1][0]{$[F(l,0)-F(l,\tau)]J/J_z$}
\psfrag{x1}[][][1][0]{$J_z/J=0.1$}
\psfrag{x2}[][][1][0]{$J_z/J=0.2$}
\psfrag{x3}[][][1][0]{$J_z/J=0.4$}
\includegraphics[width=8cm]{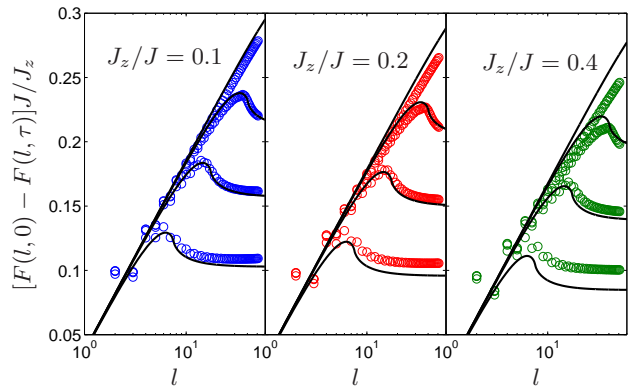}
\caption{(Color online) The fluctuation of $S_z$ in a box of length $l$ is plotted for $J_z/J=0.1$ (left panel), 0.2
  (center panel) and 0.4 (right panel) for $J\tau=3.8$, 8.7, 20.2 and 71.2 from bottom to top.  Points are iTEBD results
  and lines are fits using the bosonization result Eq.~\eqref{eq:blockfluct}.  The only global fitting parameter in each
  panel is determined as $R_0=0.5622$ (left), 0.6275 (middle) and 0.7446 (right) in units of the lattice constant.}
\label{fluctbox}
\end{figure}

The most obvious quantity to start with is the heating, i.e., the energy pumped into the system in excess of the final
ground state energy.  This is found to be \cite{doraquench,EPAPS}
\begin{gather}
\langle H \rangle=E_{gs}\left[1-
\left(\frac{\tau_0}{\tau}\right)^2
\ln\left(1+\left(\frac{\tau}{\tau_0}\right)^2\right)
\right]\;.
\label{excessenergy}
\end{gather}
Here we introduced the microscopic time scale, $\tau_0 \equiv R_0/2 v$, and $E_{gs}=-Lg_2^2/4\pi vR_0^2$ is the
adiabatic ground state energy shift to lowest order in $g_2$, with $L$ the system size.
The heating in the near-adiabatic limit in 1D gapless systems has been addressed in Refs.~\cite{pellegrini,bernier},
where non-universal behaviors were reported. The universal $\ln(\tau)/\tau^2$ heating seen in Eq.~\eqref{excessenergy}
was mentioned previously in Ref.~\cite{polkovnikovnatphys}.
In Fig. \ref{heating}, we compare Eq. \eqref{excessenergy} to the numerical result, using $R_0=0.5622$ as the only free
parameter, what we obtain from the block fluctuations (Fig. \ref{fluctbox}).

The fluctuations of $S_z$ in a given box with size $l$ (in units of the original lattice constant) are characterized by the quantity
\begin{gather}
F(l,t)=\left\langle\left[\int\limits_0^l\left( S_z(x,t)-\langle S_z(x,t)\rangle\right)dx \right]^2\right\rangle,
\end{gather}
where $S_z(x)=-\partial_x\phi(x)/\pi$
and
\begin{gather}
\phi(x)=\sum_{q\neq 0}\sqrt\frac{\pi}{2|q|L}\left(e^{iqx-\alpha|q|/2}b_q+h.c.\right).
\label{phi}
\end{gather}
In the fermionized picture, this corresponds to density fluctuations in a box, which was shown to scale identically
to the entanglement in equilibrium \cite{song}.
After bosonization, it is given by
\begin{gather}
F(l,t)=\frac{1}{\pi^2}\left\langle \left[\phi(l,t)-\phi(0,t)\right]^2\right\rangle=\nonumber\\
=\sum_{q\neq 0}\frac{4}{L|q|\pi}\sin^2\left(\frac{ql}{2}\right)\left(\frac 12 +|v_q(t)|^2+\textmd{Re}[u_q(t)v_q^*(t)]\right).
\label{phiphi}
\end{gather}
Here, the constant term  (1/2) on the r.h.s gives the fluctuations of free fermions or hard core bosons as $ F(l)\sim\ln(l)$,
while the other terms, depending on the Bogoliubov
coefficients, contain information on the quench.
The dominant contribution comes from the Re$[u_q(t)v_q^*(t)]$ term to the exponent \cite{EPAPS}, which is of order $g_2/v$, while the $v_q(t)^2$ 
goes only with  $(g_2/v)^2$, and can be neglected in the perturbative
regime.
From this, we obtain for $t=\tau$
\begin{gather}
F(l,0)-F(l,\tau)=\frac{2}{\pi^2}\frac{g_2}{2v}\left( f\left(\frac{l}{2v\tau}\right)+\ln\left(\frac{l}{R_0}\right)\right),
\label{eq:blockfluct}
\end{gather}
where 
$f(y)=\frac 12 \sum_{s=\pm 1}s(y-s)\ln|y-s|$ 
and $F(l,0)=F_{J_z=0}(l)\sim\ln(l)$ accounts for the fluctuations in the initial system, which is subtracted to focus on the effect of the quench.
In the steady state, $F(l,0)-F(l,t\rightarrow\infty)=(g_2/v\pi^2)\ln\left({l}/{R_0}\right)$ becomes independent of the
quench time, and coincides with the equilibrium fluctuations.
Right at the end of the quench with $t=\tau\gg l/v$, $F(l,0)-F(l,\tau)$ takes the same value as in the steady state,
while for $v\tau\ll l$, it saturates to $({g_2}/{v\pi^2})\ln\left({\tau}/{e\tau_0}\right)$.  These analytical results
are compared to the data obtained by the iTEBD in Fig. \ref{fluctbox}. The numerical results for the fluctuations are more sensitive to truncation effects due to a finite matrix dimension $\chi$ than the other observable we calculated. To obtain unbiased data, we performed simulations up to $\chi=2000$ and extrapolated the data to $\chi\rightarrow\infty$. For each final $J_z$, the only global fitting parameter for all
$\tau$'s is the short distance cutoff, $R_0$, which is of the order of 0.5 (in units of the original lattice constant).
For small $J_z$, the agreement is excellent. Remarkably, the semi-quantitative agreement persists for $J_z$ values as
high as $J_z=0.4$.
Also, for a given $J_z$, the agreement gets better with $\tau$, i.e.  moving towards the adiabatic limit, since the
larger $\tau$, the smaller the energy pumped into the system and the more reliable the Luttinger liquid description is.

The quantities considered so far could have in principle been obtained by using adiabatic perturbation theory
\cite{adpert}, since our perturbative results capture only the lowest order correction in $J_z/J$ to the above physical
quantities.  Therefore, we now focus on the spin flip correlation function $\langle{S^+}{S^-}\rangle$, which contains the
bosonic fields in the exponent and demonstrates the non-perturbative nature of bosonization: the present approach yields
to first correction in $J_z$ to the \emph{exponent} of the spin-flip correlation function.  Perturbation theory would
only yield the lowest order correction to the whole correlator, and not to its exponent.  Thus, it requires the
non-perturbativeness of bosonization to account for the numerical data and to produce power-law correlation functions.

The most singular, staggered part of the transverse magnetization \cite{nersesyan} is given by
\begin{gather}
S^+(x)=\frac{(-1)^x}{\sqrt{2\pi\alpha}}\exp(-i\theta(x)),
\label{splus}
\end{gather}
where $\theta(x)$ is similar to Eq. \eqref{phi}, except for an extra sgn($q$) multiplier within the $q$ summation \cite{EPAPS},
 $S^+(x)$  is also the hard core boson creation operator in the continuum limit, and $\alpha$ is a short distance cutoff.

The spin flip correlation function of the XXZ model, which corresponds to the hard-core boson single particle density matrix, $G_B(x,t)=\langle S^+(x,t)S^-(0,t)\rangle$ is 
obtained  as
\begin{gather}
G_B(x,t) =\frac{(-1)^x}{2\pi\alpha}\exp\left(-\dfrac{\left\langle\left[\theta(x,t)-\theta(0,t)\right]^2\right\rangle}{2}\right),
\label{s+s-}
\end{gather}
where $\left\langle\left[\theta(x,t)-\theta(0,t)\right]^2\right\rangle$ is similar to Eq. \eqref{phiphi}, only the sign of the last term is flipped \cite{EPAPS}.

\begin{figure}[th]
\psfrag{x}[t][][1][0]{$\log_{10}x$}
\psfrag{y}[b][t][1][0]{$\log_{10}(|G_B(x,\tau)|\sqrt{x})$}
\psfrag{x1}[][][1][0]{$J_z/J=0.1$}
\psfrag{x2}[][][1][0]{$J_z/J=0.2$}
\psfrag{x3}[][][1][0]{$J_z/J=0.4$}
\includegraphics[width=8cm]{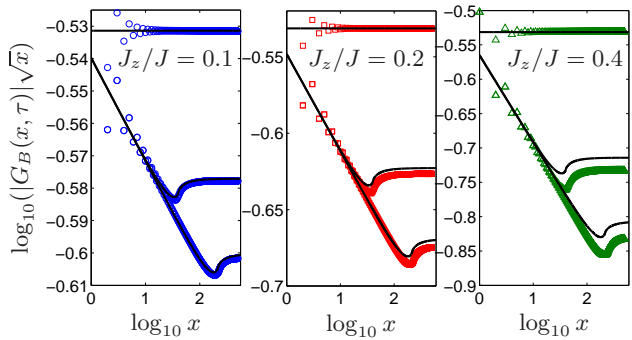}
\caption{(Color online) The spin flip correlation function is shown for $J_z/J=0.1$ (left panel), 0.2 (middle panel) and 0.4 (right panel) for $J\tau=0$, 20.2 and 108.3
 from top to bottom with $R_0=0.5622$ (from Fig. \ref{fluctbox}) from Eq. \eqref{spsm}, together with the numerical data.
The power-law exponent changes from $-\frac 12-\frac{J_z}{\pi J}$
for $x\ll v\tau$ to $-\frac 12$ for
$x\gg v\tau$,  as
$|G(B(x,\tau)|\sqrt{|x|}\approx C({R_0}/{\min[x,2v\tau/e]})^{g_2/2v}$. Results from the XY model \cite{mccoyxy} fix the prefactor of the correlation function as well, leaving $R_0$ as the only adjustable parameter.
The $\tau=0$ results correspond to that in the XX Heisenberg model \cite{mccoyxy}. At short distances, the correlator is strongly influenced by the presence of the lattice.}
\label{spsmchi}
\end{figure}

Right after the quench at $t=\tau$ and in the $|x|,v\tau\gg R_0$ limit, the spin flip correlation function reads as
\begin{gather}
G_B(x,\tau)\approx
\frac{C(-1)^x}{\sqrt{|x|}}\exp\left(-\frac{g_2}{2v}f\left(\frac{x}{2v\tau}\right)\right)\left(\frac{R_0}{x}\right)^{g_2/2v},
\label{spsm}
\end{gather}
where $C=\sqrt{e}2^{-1/3} A^{-6}$ stems from the correlator of hard core bosons on a lattice in, e.g., the XY model
($g_2=0$), $A=1.28243\dots$ is Glaisher's constant \cite{mccoyxy}.  These non-perturbative results are tested
against numerics in Fig. \ref{spsmchi}, where we fix $R_0=0.5622$ from Fig. \ref{fluctbox}.  Similarly to the previous
comparisons, the agreement is excellent and works qualitatively upto rather large $J_z$.

The short distance behaviour ($<v\tau$) in Figs. \ref{fluctbox} and \ref{spsmchi} is dominated by high energy ($>1/\tau$) modes, evolving adiabatically. The correlators thus behave identically to the adiabatic case ($\tau\rightarrow\infty$).
However, the long distance ($>v\tau$) response is dictated by low energy ($<1/\tau$) modes, feeling a sudden quench, and the observables in this range reveal the sudden 
quench behaviour ($\tau\rightarrow 0$).
We have also checked that the numerical data for time dependent correlators are also successfully described by our bosonization scheme.

 After the quench ($t\gg\tau$), Eq. \eqref{s+s-} still applies after changing $\tau$ to $t$. The momentum distribution (MD), i.e. the spatial Fourier transform
of
Eq. \eqref{s+s-}, to first order in $g_2$ behaves as
\begin{gather}
n(\tilde k,t)\sim {\tilde k}^{-1/2}\max\left(R_0\tilde k,\frac{R_0}{vt}\right)^{-g_2/2v},
\label{momdist}
\end{gather}
where $\tilde k=||k|-\pi|$.  In the steady state, it remains identical to the adiabatic expression\cite{giamarchi,nersesyan} in spite of the quench.
Had we taken a ferromagnetic coupling ($J<0$), the divergence would occur at $k=0$ as is the case normally for hard core
bosons \cite{rigol2004}.  The steady state ($t\rightarrow \infty$) response thus coincides with the equilibrium one to
first order in the exponent, irrespective of the quench time. Higher order terms, however, will modify the exponent \cite{doraquench}.
Eq. \eqref{momdist} is directly accessible experimentally using time-of-flight imaging of quenched hard core bosons.

To summarize, we have applied the Luttinger model description  for a lattice model outside the usual equilibrium purview
of this description, by deriving quantities using an out-of-equilibrium Luttinger liquid theory and comparing them to
exact numerical calculations using iDMRG/iTEBD for the XXZ chain.  Since several calculations have appeared in the literature
treating the Luttinger model in non-equilibrium situations, it is important to develop intuition for the reliability of
the Luttinger model as a description of the non-equilibrium physics of lattice models.  Our work is an important step in
that direction (cf. Ref. \cite{karrasch}).  Remarkably, even though our bosonization calculations are perturbative in $J_z$, they provide an
excellent quantitative description even for moderately large $J_z$ values.

Our work opens up a number of new questions worth pursuing in future research.  We have found bosonization to describe
well linear-quench dynamics from $J_z=0$ upto moderate values of $J_z$.  While this is indicative of the broad
applicability of bosonization out of equilibrium when starting from an initial ground state, it might also be fruitful
to explore similar issues for other non-equilibrium situations.  In particular, one might wonder if the Luttinger model
is quantitatively useful for instantaneous quenches involving large changes in $J_z$ beyond the observables considered in Ref. \cite{karrasch}, or for cases where
the initial state is not a ground state.  In general, it is not well-understood which non-equilibrium situation might
make which type of irrelevant or marginal operators important.

\begin{acknowledgments}
This research has been  supported by the Hungarian Scientific
Research Funds Nos.  K72613, K73361,
K101244, CNK80991,
T\'{A}MOP-4.2.1/B-09/1/KMR-2010-0002 and by the ERC Grant Nr. ERC-259374-Sylo.
\end{acknowledgments}

\bibliographystyle{apsrev}
\bibliography{wboson}

\newpage

\appendix*

\begin{widetext}

\section{EPAPS supplementary material}

Here we detail and explain some technical aspects of the calculation, presented in the main text.
The Jordan-Wigner transformation reads as \cite{giamarchi}
\begin{gather}
S_l^+=\exp\left(i\pi\sum_{m<l}n_m\right)c_l^+,\hspace*{3mm} S_l^-=\exp\left(i\pi\sum_{m<l}n_m\right)c_l, \hspace*{3mm} S_l^z=n_l-\frac{1}{2},\hspace*{3mm}n_l=c_l^+c_l
\end{gather}
 where the $c$'s are fermionic operators.

In terms of the time dependent Bogoliubov coefficients, the heating reads as
\begin{gather}
\langle H \rangle=2\sum_{q\neq0}
\omega(q)|v_q(t)|^2+g(q,t)\textmd{Re}[u_q(t)v_q^*(t)]=E_{gs}\left[1-
\left(\frac{\tau_0}{\tau}\right)^2
\ln\left(1+\left(\frac{\tau}{\tau_0}\right)^2\right)
\right]\;,
\end{gather}
where both terms contribute in $(g_2/v)^2$ order.

For a linear quench and $t\geq \tau$,
\begin{gather}
v(q,t)=\frac{ig_2(q)|q|}{4\omega^2(q)\tau}\left[\exp(i\omega(q)(t-2\tau))-\exp(i\omega(q)t)+2i\omega(q)\tau\exp(-i\omega(q)t)\right]
\end{gather}
and 
\begin{gather}
u(q,t)=\exp(-i\omega(q)t)
\end{gather}
to lowest order in $g_2$. Consequently, 
\begin{gather}
\textmd{Re}[u_q(t)v_q^*(t)]=-\frac{g_2(q)}{2v}\left(1+\frac{\sin(2v|q|(t-\tau))-\sin(2v|q|t)}{2v|q|\tau}\right)
\label{reuv}
\end{gather}
to lowest order in $g_2/v$. In the steady state ($t\rightarrow \infty$), the trigonometric functions average to zero and only the $-g_2(q)/2v$ factor determines the response.
The adiabatic limit, $\tau\rightarrow\infty$ leads to identical result, therefore the steady state behaves as the adiabatic limit, as if no quench occured to the system.

The bosonic field, $\theta(x)$ is defined as
\begin{gather}
\theta(x)=\sum_{q\neq 0}\sqrt\frac{\pi}{2|q|L}\textmd{sgn}(q)\left(e^{iqx-\alpha|q|/2}b_q+h.c.\right),
\end{gather}
ans its autocorrelation function reads as
\begin{gather}
\left\langle \left[\theta(x,t)-\theta(0,t)\right]^2\right\rangle =\sum_{q\neq 0}\frac{4\pi}{L|q|}\sin^2\left(\frac{qx}{2}\right)
\left(\frac 12 +|v_q(t)|^2-\textmd{Re}[u_q(t)v_q^*(t)]\right).
\end{gather}

 A quantum state on a chain of length $L$ can  be written in the following MPS form:
\begin{equation}
|\Psi \rangle = \sum_{j_1, \ldots, j_L} A^{[1]j_1}A^{[2]j_2} \ldots A^{[L]j_L} | j_1, \ldots ,j_L \rangle,  \label{eq:mps}
\end{equation}
where $A^{[n]j_n}$ are $\chi_n \times \chi_{n+1}$ matrices, and $|j_n\rangle$
represent  local states at site $n$. The  matrices at the boundary (i.e., $n=1$ and $L$) are vectors because the outer index is zero dimensional. The MPS representation is
efficient in one dimensional systems because it exploits the fact that the ground-state wave
functions are only slightly entangled \cite{hastings-2007}.

\end{widetext}
\end{document}